\documentstyle [11pt] {article}
\catcode`\@11
\def\section{
    \setcounter{equation}{0}
\@startsection {section}{1}{\z@}{-3.5ex plus -1ex minus -.2ex}
{2.3ex plus .2ex}{\large\bf}
}
\catcode`\@13
\newcommand{\nummer}[1]{\hskip 12 true cm #1 \par}
\newcommand{\netnum}[1]{\vspace{-14pt}\hskip 12 true cm #1 \par}
\newcommand{\monat}[1]{\hskip 12 true cm #1
                       \par \vspace*{1 cm}}
\newcommand{\titel}[1]{{\renewcommand{\thefootnote}{\fnsymbol{footnote}}
                       \Large\bf\vskip 0 true cm
                       \begin{center}#1\end{center}
                       \setcounter{footnote}{0}}
                       \normalsize\vskip 1.2 true cm}
\newcommand{\autor}[1]{{
                       \renewcommand{\thefootnote}{\arabic{footnote}}
                       \begin{center} {\large #1 }\end{center}}
                       \setcounter{footnote}{0}}
\newcommand{\adresse}[1]{\vspace*{-1.1 true cm}\begin{center} {\it #1 }
                         \end{center}
                         \vskip 0.5cm}

\newcommand{\bye}{\end{document}}
\newcommand{\be}{\begin{equation}}
\newcommand{\ee}{\end{equation}}
\newcommand{\bes}{\begin{eqnarray}}
\newcommand{\ees}{\end{eqnarray}}
\newcommand{\ema}{\end {array} \right)}
\newcommand{\nin}{\kern 0.1 em \in\kern -0.80em /}
\newcommand{\pslash}{\kern 0.1 em p\kern -0.45em /}
\newcommand{\dslash}{\kern 0.1 em \partial\kern -0.55em /}
\newcommand{\nabsla}{\kern 0.1 em \nabla\kern -0.70em /}
\newcommand{\sla}[1]{\kern 0.1 em #1\kern -0.55em /}
\newcommand{\nsubset}{\kern 0.1 em \subset\kern -0.80em /}

\newcommand{\lra}{\longrightarrow}

\newcommand{\R}{{I\kern -0.22em R\kern 0.30em}}
\newcommand{\N}{{I\kern -0.22em N\kern 0.30em}}
\newcommand{\C}{\mbox{\kern 0.20em \raisebox{0.09ex}{\rule{0.08ex}{1.22ex}}
                \kern -0.60em C\kern 0.30em}}
\newcommand{\Z}{{\sf Z\kern -0.40em Z\kern 0.30em}}

\begin{document}
\newcommand{\cA}{{\cal A}}
\newcommand{\cB}{{\cal B}(\cH) }
\newcommand{\cD}{{\cal D}}
\newcommand{\cH}{{\cal H} }
\newcommand{\cHA}{{\cal H}_\pi }
\newcommand{\cJ}{{\cal J} }
\newcommand{\vG}{\mbox{{\bf G}} }
\newcommand{\vg}{\mbox{{\bf g}} }
\newcommand{\cM}{{\cal M} }
\newcommand{\cN}{{\cal N} }
\newcommand{\Qb}{{\overline Q}}
\newcommand{\Qe}{\varepsilon Q}
\newcommand{\Db}{{\overline D}}
\newcommand{\parb}{{\overline \partial}}
\newcommand{\dslab}{{\overline {\dslash}}}
\newcommand{\Qbe}{{\overline \varepsilon}{\overline Q}}
\newcommand{\ia}{\alpha}
\newcommand{\ib}{\beta}
\newcommand{\ic}{\gamma}
\newcommand{\id}{\delta}
\newcommand{\iad}{{\dot{\alpha}}}
\newcommand{\ibd}{{\dot{\beta}}}
\newcommand{\icd}{{\dot{\gamma}}}
\newcommand{\idd}{{\dot{\delta}}}
\newcommand{\Th}{\theta}
\newcommand{\Thb}{{\overline \theta}}
\newcommand{\psib}{{\overline \psi}}
\newcommand{\chib}{{\overline \chi}}
\newcommand{\sigb}{{\overline \sigma}}
\newcommand{\etab}{{\overline \eta}}
\newcommand{\Phib}{{\overline \Phi}}
\newcommand{\ab}{{\overline{a}}}
\newcommand{\bb}{{\overline{b}}}
\newcommand{\cb}{{\overline{c}}}
\newcommand{\fb}{{\overline{f}}}
\newcommand{\gb}{{\overline{g}}}
\newcommand{\hb}{{\overline{h}}}
\newcommand{\vb}{{\overline{v}}}
\newcommand{\wb}{{\overline{w}}}
\newcommand{\zb}{{\overline{z}}}
\newcommand{\ve}{\varepsilon}
\newcommand{\veb}{{\overline \varepsilon}}
\newcommand{\pid}{\pi_\cD}
\newcommand{\OA}{\Omega \cA }
\newcommand{\ODA}{\Omega_\cD \cA}
\newcommand{\CLA}{{Cl}_\cD \cA}
\newcommand{\OD}[2]{\Omega_\cD^{(#1 , #2)}\!\cA}
\newcommand{\CL}[2]{{Cl}_\cD^{(#1 , #2)}\!\cA}
\newcommand{\JD}[2]{{\cal J}_\cD^{(#1 , #2)}}
\newcommand{\sd}[2]{\sigma_\cD^{(#1 , #2)}}
\begin{titlepage}
\nummer{CPT-96/P.3330}
\netnum{MZ-TH/96-07}
\netnum{hep-th/9604146}
\monat{April 23, 1996}
\titel{Supersymmetry and Noncommutative Geometry\footnote{
{\rm Work supported in part by the PROCOPE agreement between the 
University of Aix-Marseille and the Johannes Gutenberg-Universit\"at of Mainz.
}}}
\autor{W.~Kalau\footnote{e-mail: kalau{\char'100}cpt.univ-mrs.fr; 
author supported by DFG}}
\setcounter{footnote}{1}
\adresse{CNRS-CPT Luminy\\
Case 907\\
F-13288 Marseille, Cedex 9} 
\autor{M.~Walze\footnote{e-mail: walze{\char'100}vipmza.physik.uni-mainz.de}}
\adresse{Johannes Gutenberg Universit\"at\\
Institut f\"ur Physik\\
D-55099 Mainz}
\begin{abstract}
The purpose of this article is to apply the concept of the spectral triple, 
the starting point for the analysis of noncommutative spaces in the
sense of A.~Connes, to the case where the algebra $\cA$ contains both bosonic
and fermionic degrees of freedom. The operator $\cD$ of the spectral triple
under consideration
is the square root of the Dirac operator und thus the forms of the generalized 
differential algebra constructed out of the spectral triple are in a 
representation of the Lorentz group with integer spin if the
form degree is even and they are in a representation with half-integer spin
if the form degree is odd. However, we find that the 2-forms, obtained by
squaring the connection, contains exactly the components of the vector
multiplet representation of the supersymmetry algebra. This allows to
construct an action for supersymmetric Yang-Mills theory in the framework
of noncommutative geometry.
\end{abstract}
\end{titlepage}
%18-apr-1996
\section{Introduction}
In the last years it has turned out that noncommutative geometry \cite{cobuch}
offers a powerful mathematical framework for the study of fundamental 
interactions in physics. The construction of models for the electroweak and
strong interaction in terms of noncommutative geometry, i.e. the Connes-Lott
models \cite{colo,cobuch} and the model of the Marseille-Mainz group 
\cite{cev,hps1,hps2}, has led to a new qualitative insight into the
spontaneous symmetry breaking mechanism of the Standard Model. 

The basic idea of noncommutative geometry is to generalize geometric concepts
such that they can be applied to more general situations where it is
meaningless to consider e.g. points connected by arcs. This allows to relax
the physical notion of space-time such that our classical space-time may 
emerge as a ``classical-limit" of a more general noncommutative space-time.
This idea has been followed in \cite{dofrero} where uncertainty-relations for
space-time variables were implemented and their consequences were studied. 
A different approach to utilize this more general framework is to consider
a sequence of finite dimensional algebras which approximate the algebra
of functions on a classical
manifold, like the fuzzy sphere \cite{ma}, and study a quantum field
theory on such geometries \cite{mabu,groma,klim}. Here the noncommutativity of 
the geometries serves as a regulator for the field theory. 

However, the novel feature of these approaches to derive the Standard Model 
in the framework of noncommutative geometry
is not a generalization of space-time itself. In those 
models a discrete space, i.e. a space consisting of two points, is added 
to a conventional space-time.
This effects the internal symmetries of the theory such that the Higgs
becomes a part of a generalized gauge potential. It has turned out that
the Connes-Lott models and their successors \cite{co} based on real spectral 
triples do not only lead to qualitative restrictions compared to
conventional Yang-Mills Higgs models, but also serve numerical relations
for the Higgs mass and top-mass \cite{iokasch}.

We take this as a motivation to explore the concept of the spectral triple, the 
basic input data for Connes-Lott models, in a more general context. In
Connes-Lott models the notion of space-time was generalized to incorporate the
symmetry breaking mechanism of internal symmetries. In this article
we will analyse a spectral triple of a supermanifold, i.e. of a generalization
of space-time which includes fermionic degrees of freedom. This leads to a
generalization of space-time symmetry, namely to supersymmetry. The generalized
differential algebra, which is constructed out of the spectral triple, is
used to derive an action for $N=1$ supersymmetric Yang-Mills theory although
it is not the usual superdifferential algebra \cite{WB}.
The basic difference of the generalized differential algebra to the 
conventional superdifferential algebra \cite{WB} is the absence of space-time 
differentials, i.e. the absence of space-time or vectorial 1-forms. Thus the
differential algebra is generated only by spinorial 1-forms and the differential
splits into a holomorphic and an anti-holomorphic part, which are in 
$({1\over 2}, 0)$, the left-handed spin ${1\over 2}$ representation, resp.~in
$(0,{1\over 2})$, the right-handed spin ${1\over 2}$ representation of the
Spin group. As in usual Yang-Mills theory, the action is obtained by squaring
the curvature, which itself is the square of a covariant derivative.

There are several articles in the literature which deal with various
aspects of supersymmetry and noncommutative geometry. For example in
\cite{Jaffe} nocommutative geometry is applied to supersymmetric constructive
field theory. Furthermore let us just mention those articles which have a 
 direct relation to models for the electroweak interaction.
For the Marseille-Mainz model a $\Z_2$ graded structure plays a fundamental
role in the sense that this model is based on a supergroup. However, the
supergroup is not a symmetry group and therefore supersymmetry is not 
realized \cite{hps2}.

The relation of supersymmetry and Connes-Lott models was investigated by
A.~Chamseddine \cite{cha} who 
explores the possibilities of arranging the elements of spectral triples of
Connes-Lott models such the resulting model is supersymmetric. Note, however,
that although we also use the concept of spectral triples our work is
essentially different from \cite{cha}.
 
In the next section we briefly recall the definition of the spectral triple
which allows us to indicate the starting point of our construction. 
Furthermore we introduce the commutative $*$-algebra $\cA$ of the 
spectral triple which is a modified algebra of superfields. The construction 
of the spectral triple is completed by specifying the representation of $\cA$
 on a Hilbert-space $\cH$ and by defining the selfadjoint unbounded 
operator $\cD$. In section 3, 
after a brief outline of the general procedure, we start the construction
of the generalized differential algebra with the definition of the generalized
Clifford-algebra. Its holomorphic structure and the relation to supersymmetry
is discussed. The construction of the generalized differential algebra is
completed in section 4. A supersymmetric invariant inner product is constructed
in section 5.
Section 6 contains our derivation of 
supersymmetric Yang-Mills theory. This article ends with some concluding
remarks in section 7.
%18-apr-1996
\section{The spectral triple}
The basic object in noncommutative geometry defining the geometrical 
framework is the spectral triple $(\cA , \cH, \cD )$ \cite{co,cobuch}. $\cA$, 
the first element of this triple, is an associative $*$-algebra of bounded 
operators with a unit in a Hilbert-space $\cH$, the second element of the 
spectral triple. The last element, $\cD$, is a selfadjoint unbounded operator 
in $\cH$ such that
\begin{itemize}
\item[i.] $\cD$ has a compact inverse (modulo a finite dimensional kernel)
\item[ii.] $[\cD , a]=\cD a - a\cD$ is bounded  for any $a\in\cA$.
\end{itemize}
Frequently the last two objects $(\cH ,\cD)$ are called a K-cycle over $\cA$.
These three elements together encode all geometric information of a space
as spectral data. For example, it is possible to construct a differential
algebra for this space, where the operator $\cD$ defines the differential. 
This is the starting point for Yang-Mills theory in noncommutative geometry
\cite{cobuch} (see, e.g., \cite{kabuch,vabo} for a review). 
We should mention that we gave only the definition of 
spectral triples of compact spaces. However, it is also possible
to define spectral triples for spaces which are only locally compact \cite{co}.

The spectral triple describing the geometry of a compact
spin-manifold $\cM$ is given by\linebreak 
$(C^\infty(\cM), L_2(S), D)$, where 
$C^\infty(\cM)$ are the smooth functions on $\cM$, $L_2(S)$ is the Hilbert-space
of square-integrable spin-sections and $D$ is the usual Dirac-operator
\cite{co,cobuch}. The differential algebra derived from this triple is the
de Rham algebra of differential forms on $\cM$.

Let us express the operator $\cD$ of this example, the
Dirac-operator, 
in a somewhat different terms which refers to (symmetry) transformation
of $\cM$. Thus we think of the Dirac-operator $D=\gamma^\mu\nabla_\mu$ 
as a composition of two kinds of objects:
\begin{itemize}
\item[i.] the generators of parallel displacement or covariant derivative
$\nabla_\mu$,
\item[ii.] the generators of the Clifford-algebra corresponding to the vector
space of generators of parallel displacement.
\end{itemize}
There is a well known generalization of this Lie-algebra of parallel 
displacement in physics: the supersymmetry
(for the rest of the article we restrict ourselves to the case in 
which the manifold $\cM$ is flat), which is generated by $Q$ and $\Qb$.
The fundamental commutation relation is
\be
[\Qe , \Qbe ] = 2i \ve \sigma^\mu \veb \partial_\mu \label{susydef}
\ee
where $\ve$  is a constant anticommuting chiral spinor and $\veb$ is a
antichiral spinor, related to $\ve$ by charge conjugation, i.e. $\ve$
and $\veb$ together form a Majorana-spinor. This implies
that in 4 dimensions an Euclidean space-time metric is excluded.
However, the noncommutative analogue of an integral, the Dixmier trace, is
defined only on Euclidean space. On the other hand, the construction of the
generalized differential algebra does not refer to the signature of
space-time. Furthermore, the special structure
of the Hilbert-space, which will be defined below, allows to
define an inner product on the generalized Clifford-algebra which induces an inner product on the generalized differential forms. This leads to a 
supersymmetric invariant action which is defined without using the Dixmier-trace.

The purpose of this article is to encode the generalization of space-time
in the sense of eq.(\ref{susydef}) in the spectral triple. Therefore we have 
to extend
the algebra of (bosonic) functions, $C^\infty(\cM)$ by fermionic quantities.
Thus we have to include the spinors as anticommuting, i.e. as Grassmann-odd,
objects in the algebra. In order to maintain
the regularity of the algebra we restrict ourselves to the dense subspace
of smooth spinors $\Gamma(S)\subset L_2(S)$. 
Furthermore it is useful to split
$\Gamma(S)$ into its irreducible parts of the Lorentz-group:
\be
\begin{array}{rcl}
\Gamma(S) &=& \Gamma(S_+) \oplus \Gamma(S_-)\\
  & & \\
\Psi &=& (\psi_\ia, \chib^\iad)
\end{array}
\ee
The indices $\ia\in\{1,2\}$ and $\iad\in\{1,2\}$ can be raised and lowered
with the antisymmetric tensors $\ve^{\ia\ib}$ and $\ve_{\iad\ibd}$. We use
the convention of \cite{WB}, i.e.
\be
\ve_{21}= \ve^{12} = 1
\ee
for both $\ve$-tensors with dotted and undotted indices.

The multiplication rules of
spinors are most conveniently described with the help of a constant 
anticommuting Majorana spinor $\Th$. 
Thus the algebra $\cA_0$ is generated by elements $g_0$ of the form
\be
g_0= f + \Th^\ia\psi_\ia + \Thb_\iad\chib^\iad\;\;\; ,\;\;\; 
f\in C^\infty(\cM),\; (\psi_\ia ,\chib^\iad )\in\Gamma(S)
\label{A0-def}
\ee
is the usual algebra of superfields \cite{WB}. A general element $a_0\in\cA_0$
can be expanded in powers of $(\Th, \Thb)$ as follows
\be
a_0= a_1 +\Th^\ia{a_2}_\ia + \Thb_\iad{a_3}^\iad
 +\Th^\ia\Thb_\iad{a_4}^\iad_\ia+ \Th\Th{a_5} + \Thb\Thb{a_6}+
 +\Thb\Thb\Th^\ia{a_7}_\ia + \Th\Th\Thb_\iad{a_8}^\iad +\Th\Th\Thb\Thb{a_9}\; .
\ee

The $*$-operation is
defined on the generators $g_0$ as 
complex conjugation on functions and charge conjugation on
spinors. This definition extends uniquely to the whole algebra $\cA_0$.

However, this algebra is not
well suited for our purpose as will become clear when we compute the 
generalized differential algebra. Therefore we enlarge the algebra by taking
spinor doublets as generators of the algebra $\cA$, i.e., we define 
$\cA$ to
be the algebra which is generated by elements $g$ of the following form
\be
g= f + \Th^\ia\psi_\ia\otimes v +\Thb_\iad\chib^\iad\otimes \wb ,\;\;\;
f\in C^\infty(\cM),\; (\psi_\ia, \chib^\iad)\in\Gamma(S) ,\;
v,\wb\in \C^2\; .\label{A-def}
\ee
For a generator $g$ as in eq.(\ref{A-def}) the $*$-operation is defined to
be
\be
g^*= \fb + \Th^\ia\chi_\ia\otimes w +\Thb_\iad\psib^\iad\otimes \vb\; .
\ee
The multiplication for elements in $\C^2$ is just the totally symmetrized
tensor-multiplication.
Thus a general element $a\in\cA$ can be expanded in powers of $(\Th,\Thb)$ as
follows
\be
\begin{array}{rl}
a= &a_1 +\Th^\ia{a_2}_\ia\otimes v^{(2)} +
\Thb_\iad{a_3}^\iad \otimes v^{(3)}\\
 & \\ 
 &+\Th^\ia\Thb_\iad{a_4}^\iad_\ia\otimes (v^{(4)}_1\otimes_s v^{(4)}_2) +
\Th\Th{a_5}\otimes (v^{(5)}_1\otimes_s v^{(5)}_2) +
\Thb\Thb{a_6}\otimes (v^{(6)}_1\otimes_s v^{(6)}_2) \\
 & \\
 &+\Thb\Thb\Th^\ia{a_7}_\ia\otimes (v^{(7)}_1\otimes_s v^{(7)}_2 \otimes_s
 v^{(7)}_3)+
\Th\Th\Thb_\iad{a_8}^\iad\otimes (v^{(8)}_1\otimes_s v^{(8)}_2\otimes_s
 v^{(8)}_3) \\
 & \\
 &+\Th\Th\Thb\Thb{a_9}\otimes (v^{(9)}_1\otimes_s v^{(9)}_2\otimes_s v^{(9)}_3
\otimes_s v^{(9)}_4)\; . 
\end{array} \label{theta-exp}
\ee
There is no direct definition of supersymmetry generators on $\cA$ which
could be obtained as generalization of the supersymmetry generators on
$\cA_0$ which are defined as follows:
\be
Q_\ia = \partial_\ia - i \dslash_{\ia\iad}\Thb^\iad\;\;\;\; ,\;\;\;\;
\Qb_\iad = -\parb_\iad + i \Th^\ia\dslash_{\ia\iad}\; .\label{Q-def}
\ee
Here $\partial_\ia ={\partial\over \partial \Th^\ia}$, 
$\parb_\iad ={\partial\over \partial \Thb^\iad}$, denote the derivatives 
with respect to $\Th^\ia$ and $\Thb_\iad$ and \linebreak
$\dslash_{\ia\iad}=\sigma_{\ia\iad}^\mu\partial_\mu$, 
where $\sigma^i= -\sigb^i$, $i=1,2,3$ denote the Pauli matrices and 
$\sigma^0=\sigb^0=1_{2\times2}$.

However, for any fixed $v\in\C^2$ with
$\vb=v$ there is an embedding 
\be
i_v : \cA_0 \lra \cA
\ee
which is defined on the generators of $\cA_0$ as
\be
i_v( f + \Th^\ia\psi_\ia + \Thb_\iad\chib^\iad)
=( f + \Th^\ia\psi_\ia\otimes v + \Thb_\iad\chib^\iad\otimes v)
\;\;\; ,\;\;\; 
\forall f\in C^\infty(\cM),\; 
\forall (\psi_\ia ,\chib^\iad) \in\Gamma(S)\label{i_v}
\ee
Thus this allows us to define the supersymmetry generators on the subalgebra 
$\cA_v = i_v(\cA_0)$ of $\cA$ as
\be
Q_\ia^{(v)} = i_v Q_\ia i_v^{-1}\;\;\;\; , \;\;\;\;
\Qb_\iad^{(v)} = i_v \Qb_\iad i_v^{-1}\; .\label{Qv-def}
\ee
Explicitly these generators read
\be
Q_\ia^{(v)} = \partial_\ia\otimes v^* - i \dslash_{\ia\iad}\Thb^\iad\otimes v
\;\;\;\; ,\;\;\;\;
\Qb_\iad^{(v)} = -\parb_\iad\otimes v^* + i \Th^\ia\dslash_{\ia\iad}\otimes v
\; ,\label{Qv-def1}
\ee
where $v^*$ denotes the dual vector of $v$, i.e. $v^*(v)=1$. However, the
action of $v^*$ on higher powers of $v$ is defined to be
\be
v^*(v^n)=v^*(v\otimes_s \cdots \otimes_s v) = v^{n-1}\;.
\ee 
This definition follows directly from eq.(\ref{Qv-def}). The action of $v^*$
on higher powers of $v$ differs from the action of a derivative and therefore
there is no direct extension of this definition to symmetric tensor products
of arbitrary vectors. On the other hand, if $v^*$ would act like a derivative
on tensor products, the operators defined in eq.(\ref{Qv-def1}) would not
generate a supersymmetry algebra.

We now turn to the next element of the spectral triple, the space $\cH$,
which carries a representation of $\cA$. A representation space $\cHA$
can be constructed out of the algebra $\cA_0$ generated by elements of
the form as in eq.(\ref{A0-def}) as follows:
\be
\cHA = \cA_0\otimes\Z = \bigoplus_{n\in\Z} \cA_n
\ee
where we have defined
\be
\cA_n = \cA_0\otimes n \;\;\; , \;\;\; \forall n\in\Z\; .
\ee
Thus $\cHA$ is the $\Z$-fold copy of $\cA_0$.
We will call the index $n$ the S number, i.e. for all elements 
$\Phi_n \in\cA_n$ it is
\be
S(\Phi_n) = n\; .
\ee
The inner product $(\cdot, \cdot)$ on this space we define for any 
$\Psi_k=\Psi\otimes k\in\cA_k$, $\Phi_l=\Phi\otimes l\in\cA_l$ as
\be
(\Psi_k , \Phi_l) = \delta_{k+l,0}\int_\cM {dV} (\Psi^*\Phi)|_{\Th\Th\Thb\Thb}
\label{in-prod}
\ee
where $|_{\Th\Th\Thb\Thb}$ denotes the projection onto the 
$\Th\Th\Thb\Thb$-component of the $\Th$, $\Thb$ expansion of the superfields.
Thus it is the usual indefinite inner product on the algebra of superfields
$\cA_0$ multiplied by an indefinite inner product on $\Z$. Note, that the
supersymmetry generators, as defined in eq.(\ref{Q-def}), also are well
defined on $\cHA$ and that the inner-product, defined in eq.(\ref{in-prod}),
is invariant under supersymmetry transformations.

For the definition of a representation of $\cA$ on $\cHA$ we have to
introduce two operators $S_+$ and $S_-$ which act on elements 
$\Psi\otimes k \in \cHA$ as follows
\be
\begin{array}{rcl}
S_+(\Psi\otimes k) &=& \Psi\otimes(k+1)\\
 & & \\
S_-(\Psi\otimes k) &=& \Psi\otimes(k-1)
\end{array}\;\;\;\; , \;\; \forall k \in \Z \; .
\ee
These $S_+$ and $S_-$  are selfadjoint and $S_+S_-=S_-S_+=1$.
With these operators we can define a representation $\pi(a)$ on $\cHA$ for
all $a \in\cA$ with an $\Th$, $\Thb$-expansion as in eq(\ref{theta-exp})
as follows
\be
\begin{array}{rl}
\pi(a) = &a_1 +\Th^\ia{a_2}_\ia\otimes V_1^{(2)} +
\Thb_\iad{a_3}^\iad \otimes V_1^{(3)}\\
 & \\ 
 &+\Th^\ia\Thb_\iad{a_4}^\iad_\ia\otimes V^{(4)}_1V^{(4)}_2 +
\Th\Th{a_5}\otimes V^{(5)}_1V^{(5)}_2 +
\Thb\Thb{a_6}\otimes V^{(6)}_1V^{(6)}_2 \\
 & \\
 &+\Thb\Thb\Th^\ia{a_7}_\ia\otimes V^{(7)}_1 V^{(7)}_2 V^{(7)}_3+
\Th\Th\Thb_\iad{a_8}^\iad\otimes V^{(8)}_1 V^{(8)}_2 V^{(8)}_3 \\
 & \\
 &+\Th\Th\Thb\Thb{a_9}\otimes V^{(9)}_1 V^{(9)}_2 V^{(9)}_3 V^{(9)}_4\; . 
\end{array} \label{Ap-def}
\ee
where we used the notation
\be
V_i^{(j)}= {(v_i^{(j)})}_+ S_+  + {(v_i^{(j)})}_- S_-\;\;\; ,\;\;\;
v_i^{(j)}=({(v_i^{(j)})}_+ , {(v_i^{(j)})}_-)\in\C^2\; .
\ee
{}From eq.(\ref{Ap-def}) we can read off the range of the $S$-numbers of
the components in the $\Th$, $\Thb$ expansion 
\be 
S(a_k) \in 
\{-n, -n+2, \cdots ,n-2, n\}
\ee
where $n\leq 4$ is the power of $(\Th ,\Thb)$ at which the component appears.

Due to the fact that $\cA_0$ has a unit element $1$, there is an invariant
subspace $\cH_\cA \subset\cHA$ which is generated by $\cA$
\be
\cH_\cA = \cA (1\otimes 0) = \cA |0\!>\; .\label{H_a-def}
\ee
This allows us to define an (indefinite) inner product on $\cA$ as
\be
< a , b >  = ( (a |0\!>) , (b |0\!>) ) = <\!0| a^* b |0\!>
\;\;\; , \;\;\; \forall a,b\in\cA\; .
\label{A-norm}
\ee
Note, that this inner product is degenerate for the components $a_7$, $a_8$
and $a_9$ with the following $S$-numbers
\be
S(a_7) = S(a_8) = \pm 3\;\;\; ,\;\;\; S(a_9) = \pm 2,\pm 4
\ee

Eq.(\ref{A-norm}) induces also an inner product on $\cA_v$ which 
depends on $v=(v_+, v_-)\in \C^2$ and can be completely degenerate
\be
< a_v , b_v > = < \pi \circ i_v( a_0), \pi \circ i_v (b_0) > =
6 v_+^2 v_-^2 (a_0,b_0)\;\;\; ,\;\;\; \forall a_0, b_0 \in \cA_0\; .
\ee
Thus we are led to require
\be
v_+ \neq 0 \;\;\; ,\;\;\; v_- \neq 0\; .
\ee

We now turn to the last element of the spectral triple, the unbounded 
selfadjoint operator $\cD$. We construct this operator out of the
two operators $D_\ia$ and $\Db_\iad$
\be
D_\ia=\partial_\ia + i \dslash_{\ia\iad}\Thb^\iad\;\;\; , \;\;\;
\Db_\iad= -\parb_\iad - i \Th^\ia\dslash_{\ia\iad} \label{ddef}
\ee
which are associated to the supersymmetry generators as defined in 
eq.(\ref{Q-def}). By construction they enjoy the property that
all anticommutators of these operators with the supersymmetry generators
vanish and the only non-vanishing anticommutator is
\be
[ D_\ia, \Db_\iad ]_+ = -2i\dslash_{\ia\iad}\;\; .
\ee
Furthermore, we can use them to define the following subalgebras of $\cA$
\be
\begin{array}{rcl}
\cA_+ &=& \{ a\in \cA \; | \; \Db_\iad a = 0\;\}\\
  &  \\
\cA_- &=& \{ a\in \cA \; | \; D_\ia a = 0\;\}
\end{array}
\ee
The algebra $\cA$ can be generated with the two algebras $\cA_+$ and $\cA_-$,
i.e. any element $a\in\cA$ can be written as
\be
a= \sum_i a_+^{(i)}a_-^{(i)}\;\;\; ,\;\;\;\; 
a_+^{(i)}\in\cA_+,\;a_-^{(i)}\in \cA_-\;.
\ee
This fact will turn out to be important for the complex structure of the
super-Clifford algebra and it will be
very useful in the computation of $\ODA$.

We only use the two operators for the construction of $\cD$ 
since the space-time derivatives are already encoded in $D$ and $\Db$.
In other words, the operator $\cD$ is not constructed out of the full set of
operators which form a basis of the supersymmetry algebra as a vector space.
$\cD$ contains only the generating operators from which the complete algebra is
obtained via commutation relations. Thus the operator $\cD$ constructed
out of $D$ and $\Db$ has a natural interpretation as a square root of the
Dirac-operator. As a consequence the 1-forms of
the resulting differential algebra will be in the spin ${1\over 2}$
representation of the Lorentz group. 

Having fixed the derivative part of $\cD$ we still have to construct the
`Clifford algebra' part. However, since the operators $D$ and $\Db$ are odd,
i.e., they obey anticommutation relations the corresponding `Clifford algebra'
has to fulfill the following commutation relations
\be
\begin{array}{rcl}
\eta^\ia \eta^\ib - \eta^\ib \eta^\ia &=& 2i \ve^{\ia\ib} \\
                                      & &                 \\
\etab_\iad \etab_\ibd - \etab_\ibd \etab_\iad &=& 2i \ve_{\iad\ibd} \\
                                              & &                   \\
\eta^\ia \etab_\ibd - \etab_\ibd \eta^\ia &=& 0 \; ,
\end{array}
\label{defheis}
\ee
where the right hand sides are dictated by the symplectic form which defines 
the inner product on spinors. The $\eta^\ia$ and $\etab_\iad$ have to be
related by hermitean conjugation since the operator $\cD$ has to be self 
adjoint. Thus eq.(\ref{defheis}) defines a Heisenberg algebra which has a
unitary representation on $\cH_H=L_2(\C\oplus \overline{\C})$.

The total space $\cH$ of the spectral triple is the tensor product of
the representation space of $\cA$ and the representation space of the
Heisenberg algebra
\be
\cH = \cH_H \otimes \cH_\cA
\ee
and the operator $\cD$ is defined on this space as
\be
\cD= \eta^\ia\otimes D_\ia + \etab_\iad\otimes\Db^\iad\;\; .\label{DO-def}
\ee
Unless there is no risk of confusion we drop the tensor notation and
simply write $\pi(a) =a$ and $\cD= \eta^\ia D_\ia + \etab_\iad\Db^\iad$.
%18-apr-1996
\section{The universal differential envelope and the super-Clifford algebra}
Let us start this section with a brief description of the general construction
of a generalized differential algebra in noncommutative geometry \cite{cobuch} 
(for detailed reviews the reader may consult \cite{vabo,kabuch}).

The first step is to construct the universal differential envelope 
$\Omega\cA$ by associating to each element $a\in\cA$ the symbol
$\delta a$. $\Omega\cA$ is the free algebra generated by the symbols $a$,
$\delta b$, with $a,b\in\cA$, modulo the relation
\be
\delta (ab)=\delta a\, b + a\delta b\;\; .
\ee
With the definition
\be
\begin{array}{rcl}
\delta(a_0\delta a_1\cdots\delta a_k) & \;:= &
\;\delta a_0\,\delta a_1\cdots\delta a_k \\
 & & \\
\delta(\delta a_1\cdots\delta a_k) & \;:= & 0
\end{array}
\ee
$\Omega\cA$ becomes a $\N$-graded differential algebra with the odd
differential $\delta$, $\delta^2=0$. 
By defining
\be
{\delta(a)}^*=-\delta(a^*)
\ee
the $*$-operation is extended uniquely to $\Omega\cA$.

The next step is to extend the representation $\pi$ of $\cA$ to a 
representation $\pid$ of $\OA$. Since $[\cD, a]$ is bounded for any $a\in\cA$
we can define for all $k\in \N$
\be
\begin{array}{rcl}
\pid : \Omega^k\cA & \lra & \cB\\
    &      &    \\
\pid(a_0\delta a_1 \cdots \delta a_k) &=& a_0[\cD,a_1]\cdots [\cD,a_k]\; .
\end{array}
\ee
Although $\pid$ is a representation of the algebra $\OA$ it fails to be
a homomorphism of differential algebras. The trouble is that from 
\be
\pid(\omega)= 0\;\;\;,\;\;\;\omega \in \OA
\ee
it does not follow that
\be
\pid(\delta\omega)=0\; .
\ee
To obtain a differential algebra one has to identify these disturbing
elements which form a graded differential ideal $\cJ$. This ideal is given 
as \cite{cobuch}
\be
\begin{array}{rcl}
\cJ^n &=& (\ker \pid \cap \Omega^n\cA) \cup
\delta(\ker \pid \cap \Omega^{n-1}\cA)\\
     & & \\
\cJ    &=& \bigoplus_{n\in\N} \cJ^n\; .
\end{array}
\ee
Finally, the generalized differential algebra $\ODA$ is defined as the  
following quotient algebra
\be
\begin{array}{rcl}
\Omega_\cD^n\cA &=& {\Omega^n\cA\over \cJ^n}\\
                & & \\
\ODA    &=& \bigoplus_{n\in\N} \Omega_\cD^n\; .
\end{array}
\ee
However, before we start to compute $\cJ$ and $\ODA$ let us first discuss
the representation $\pid$ of $\OA$. For the spectral triple 
$(C^\infty(\cM), L_2(S), \dslash )$ the image $\pi_{\dslash}(\Omega\cA)$ is 
the Clifford-bundle $Cl(\cM)$ over $\cM$ \cite{cobuch}.
Thus in our case we call the image of $\pid$, $\CLA = \pid(\OA)$, the 
super-Clifford algebra of the spectral triple $(\cA,\cH,\cD)$. From the fact 
that $\cA$ is generated by chiral and anti-chiral superfields and
\be
[ \eta^\ia D_\ia, a] [\etab_\iad \Db^\iad, b] +
[\etab_\iad \Db^\iad, b] [ \eta^\ia D_\ia, a] = 0
\;\;\; ,\;\;\; \forall a,b\in\cA
\ee
we conclude that 
\be
\pid(\Omega^1\cA)=\CL{1}{0} \oplus \CL{0}{1}
\label{holo-cl1}
\ee
where $\CL{1}{0}$ and $\CL{0}{1}$ are linear spaces defined as 
\be
\begin{array}{rcl}
\CL{1}{0}&=&\{\pid(\sum_i a^{(i)}\delta b^{(i)}_+)=\sum_i 
 a^{(i)}\eta^\ia D_\ia b^{(i)}_+|\;
 a^{(i)}\in\cA,\; b^{(i)}_+\in\cA_+\}\\
 & & \\
\CL{0}{1}&=&
\{\pid(\sum_ia^{(i)}\delta b^{(i)}_-)=\sum_i
a^{(i)}\etab_\iad\Db^\iad b^{(i)}_-|\;
a^{(i)}\in\cA,\; b^{(i)}_-\in\cA_-\}\; .
\end{array}\label{holo-def-cl}
\ee
The algebra $\OA$ is generated by 1-forms $a\delta b \in\Omega^1\cA$
therefore the decomposition (\ref{holo-cl1}) extends to the images of
higher forms
\be
\pid(\Omega^k\cA) = \bigoplus_{l=0}^k \CL{k-l}{l}\; . \label{holo-cl}
\ee
A generic element $v\in\CL{k}{l}$ is of the form
\be
\begin{array}{rcl}
v&=&\eta^{\ia_1}\cdots\eta^{\ia_k}\etab_{\iad_1}\cdots\etab_{\iad_l}
    v_{\ia_1\cdots\ia_k}^{\iad_1\cdots\iad_l}\\
 & & \\
 &=& \eta^{\ia_1}\cdots\eta^{\ia_k}\etab_{\iad_1}\cdots\etab_{\iad_l}
\left( \sum_i a^{(i)}[D_{\ia_1}, b_1^{(i)}]\cdots[D_{\ia_k} , b_k^{(i)}]
       [\Db^{\iad_1}, c_1^{(i)}]\cdots[\Db^{\iad_l} , c_l^{(i)}]\right)
\end{array}
\ee
with $a^{(i)}\in\cA$, $b^{(i)}\in\cA_+$ and $c^{(i)}\in\cA_-$. Thus the 
elements of $\CL{k}{l}$ are tensor superfields with $k$ holomorphic
and $l$ anti-holomorphic spinor indices.

Let us now turn to the $S$-numbers of the elements in $\CLA$. The 
$\partial_\ia$, resp.~$\parb_\iad$ part of the operator $D_\ia$, 
resp.~$\Db_\iad$ shifts the coefficients of higher powers of $\Th$, $\Thb$
to lower powers and thus also the number of $S_\pm$ operators (which
coincides with the power of $\Th$, $\Thb$ for elements in $\cA$) is shifted
to lower powers of $\Th$, $\Thb$. Thus for any element 
$\omega\in \CL{k}{l}$ with a $\Th$, $\Thb$-expansion as in (\ref{Ap-def})
the range for the $S$-numbers of the coefficients of $(\Th , \Thb)^n$ is
\be
S(\omega_j) = \{-(k+l+n), -(k+l+n) +2, \cdots, k+l+n-2, k+l+n\}
\ee
where we suppressed the explicit dependence of $\eta$, $\etab$ and $S_\pm$.

Again, for any real $v\in \C^2$ there is a subalgebra of $\CLA$
on which there is a well defined action of supersymmetry. Thus there is
an extension of $i_v$ as defined in eq.(\ref{i_v}) which embeds 
$\cA^{(k,l)}_0$, i.e. tensor superfields with $k$ holomorphic and $l$ 
anti-holomorphic indices, into $\CL{k}{l}$. We define this embedding 
$i_{^{k+l}}$ on the components of the $(\Th,\Thb)$-expansion as
\be
i_{v^{k+l}}((\Th,\Thb)^n \omega_{(n)}) = 
(\Th, \Thb )^n\omega_{(n)} V^{n+k+l}\; .\label{ivn}
\ee
However, note that $(\CLA)_{v^{k+l}}=i_{v^{k+l}}(\cA^{(k,l)}_0)$ is not 
invariant under the action of $\cD$: 
\be
\begin{array}{rcl}
{[} i \eta^\ia \dslash_{\ia\iad}\Thb^\iad , (\CL{k}{l})_{v^{k+l}} ] 
&\nsubset & (\CL{k+1}{l})_{v^{k+l+1}}\\
 & & \\
{[} i \etab_\iad \dslab^{\iad\ia}\Th_\ia , (\CL{k}{l})_{v^{k+l}} ] 
&\nsubset & (\CL{k}{l+1})_{v^{k+l+1}}
\end{array}
\ee
because these parts of $\cD$  proportional to $\Th$ resp. $\Thb$
which causes a shift of components of lower powers of $(\Th, \Thb)$ to
higher powers of $(\Th ,\Thb)$ whereas the power of $V$ remains unchanged.

However, the embedding $i_{v^{k+l}}$ defined in eq.(\ref{ivn}) can be 
generalized in the following way
\be
i_{v^{k+l-2m}}((\Th,\Thb)^n \omega_{(n)}) = 
(\Th, \Thb )^n\omega_{(n)} V^{n+k+l-2m}\;\;\;\; ,\;\;\;\; 0\leq 2m \leq k+l\: .
\ee
For all $k,l\in\N$ and $2m\leq k+l$ this defines a series of subspaces 
$(\CL{k}{l})_{v^{k+l-2m}}\subset\CL{k}{l}$ which
carry a representation of the supersymmetry algebra. It is easy to check 
that these subspaces form a subalgebra of $\CLA$
\be
(\CL{k}{l})_{v^{k+l-2m}}\cdot(\CL{r}{s})_{v^{r+s-2n}}=
(\CL{k+r}{l+s})_{v^{k+r+l+s-2(m+n)}}
\ee
and we define
\be
(\CLA)_v = 
\bigoplus_{k,l\in\N}\bigoplus_{m=0}^{2m=k+l}(\CL{k}{l})_{v^{k+l-2m}}\; .
\ee
Note that it is
\be
[\cD , (\CLA)_v ] \subset (\CLA)_v \; .
\ee
%18-april-1996
\section{The generalized differential algebra $\ODA$}
Now we turn to the computation of $\cJ$ resp. $\cJ_\cD = \pid(\cJ)$.
The decomposition of $\pid (\Omega^k\cA)$ in eq.(\ref{holo-cl}) induces the 
decomposition 
\be
\cJ_\cD^k = \bigoplus_{l=0}^k \JD{k-l}{l}
\ee
and also
\be
\Omega_\cD^k\cA= \bigoplus_{l=0}^k \OD{k-l}{l}
               = \bigoplus_{l=0}^k {\CL{k-l}{l}\over \JD{k-l}{l}}\; .
\ee
Since $\pid(\cJ^1)=\{0\}$ the first non-trivial contributions to $\cJ_\cD$
appears at the level of two forms, which splits into three parts
\be
\pid(\cJ^2)= \JD{2}{0} \oplus \JD{0}{2} \oplus \JD{1}{1}\; .\label{split-J}
\ee
The first two spaces on the right hand side of eq.(\ref{split-J}), i.e. the
holomorphic and anti-holomorphic part of $\cJ_\cD^2$ are determined
by the following line of arguments:\linebreak
 for any $a,b \in \cA$ we set
$$
\begin{array}{crcl}
 &\omega &=& -\delta(ab) + a\delta b + b\delta a \in \Omega^1\cA\\
 &       & & \\
\Rightarrow &\pid(\omega)&=& 0 \; .
\end{array}
$$
Thus it is $\delta\omega \in \cJ^2$ and we compute
\be
\begin{array}{rcl}
\pid(\delta\omega) &=& 
\eta^\ia\eta^\ib ([D_\ia , a][D_\ib , b]+[D_\ia , b][D_\ib , a] ) +
\etab_\iad\etab_\ibd ([\Db^\iad , a][\Db^\ibd , b]+
[\Db^\iad , b][\Db^\ibd , a] )\\
 & & \\
 &=& 2i\ve^{\ia\ib}[D_\ia ,a][D_\ib, b]
     + 2i\ve_{\iad\ibd}[\Db^\iad ,a][\Db^\ibd, b] \; .
\end{array}
\ee
{}From this we conclude that the holomorphic part $\JD{2}{0}$ contains
all antisymmetric tensor superfields, i.e.
\be
\JD{2}{0}=\{\eta^\ia\eta^\ib w_{\ia\ib} \in \CL{2}{0} | 
w_{\ia\ib} =- w_{\ib\ia} \}\label{holo-J}
\ee 
and also for the anti-holomorphic part we find
\be
\JD{0}{2}=\{\etab_\iad\etab_\ibd w^{\iad\ibd} \in \CL{0}{2} | 
w^{\iad\ibd} =- w^{\ibd\iad} \}\; . \label{aholo-J}
\ee 
$\JD{2}{0}$ already generates the complete holomorphic part of $\cJ_\cD$,
i.e. for any $k\in\N$, $k\geq2$ it is
\be
\JD{k}{0}=\bigcup_{k=0}^{k-2} \CL{k-l-2}{0}\JD{2}{0}\CL{l}{0}\; .
\ee
To see that this holomorphic ideal in $\CL{\bullet}{0}$ is the correct ideal it
is sufficient to show that the holomorphic algebra $\Omega_\cD^h\cH$ with
\be
\Omega_\cD^h\cA =
\bigoplus_{n\in\N} \OD{n}{0} =
\bigoplus_{n\in\N} {\CL{n}{0}\over\JD{n}{0}} \label{holo-D}
\ee
is a differential algebra. The algebra defined in eq.(\ref{holo-D}) 
contains only totally symmetric tensor superfields i.e., it is
\be
\OD{k}{0}=\{\eta^{\ia_1}\cdots\eta^{\ia_k}w_{\ia_1\cdots\ia_k}\in\CL{k}{0}|
w_{\ia_1\cdots\ia_k} = {1\over k!} w_{(\ia_1 \cdots\ia_k)}\}\; ,
\ee
where $(\ia_1 \cdots \ia_k )$ denotes the sum over all permutations of the
enclosed indices.
The map $\sd{\bullet}{0}$ from the holomorphic super-Clifford algebra onto the
holomorphic differential algebra can be most conveniently defined as
\be
\begin{array}{rcl}
\sd{k}{0} & : & \CL{k}{0} \lra \OD{k}{0}\\
          &   &                          \\
\sd{k}{0}(\eta^{\ia_1}\cdots\eta^{\ia_k}w_{\ia_1\cdots\ia_k}) &=&
z^{\ia_1}\cdots z^{\ia_k}w_{\ia_1\cdots\ia_k}\; ,
\end{array}
\ee
where $(z^{\ia_i})$ denote the basis 1-forms which are complex, 
Grassmann-even, vectors with two components
, i.e., $z\in \C^2$ and
\be
z^\ia z^\ib - z^\ib z^\ia = 0\; .
\ee
The holomorphic differential $d_h$ on $\Omega_\cD^h\cA$ is a differential
of degree $(1,0)$
\be
\begin{array}{rcl}
d_h & : &  \OD{k}{0} \lra \OD{k+1}{0}\\
 & & \\
d_h(z^{\ia_1}\cdots z^{\ia_k}w_{\ia_1\cdots\ia_k}) & = & {1\over k+1}
z^{\ia_1}\cdots z^{\ia_{k+1}}\left( \sum_{l=1}^{k+1}
D_{\ia_l} w_{\ia_1\cdots\ia_{l-1}\ia_{l+1}\cdots\ia_{k+1}}\right)\; .
\end{array}
\ee
Since
\be
D_\ia D_\ib + D_\ib D_\ia = 0
\ee
it is
\be
d_h^2=0
\ee
Also it follows from this anti-commutation relation by the graded Jacobi
identity that for any 
$w_{\ia_1\cdots\ia_k}=a_0[D_{\ia_1}, a_1]\cdots[D_{\ia_k}, a_k]$ it is
\be
d_h(z^{\ia_1}\cdots z^{\ia_k}w_{\ia_1\cdots\ia_k})  = z^{\ia_0}\cdots z^{\ia_k}
[D_{\ia_0},a_0][D_{\ia_1}, a_1]\cdots[D_{\ia_k}, a_k]\; ,
\ee
which shows that $(\Omega_\cD^h\cA , d_h)$ describe correctly the pure 
holomorphic part of $\ODA$. 

The anti-holomorphic part of $\ODA$ can be obtained by analogous arguments
or, alternatively, by the fact that the holomorphic and anti-holomorphic part
are related by hermitean conjugation:
\be
{(\CL{k}{0})}^* = \CL{0}{k} \;\; \Rightarrow \;\; {(\JD{k}{0})}^* = \JD{0}{k}
\ee
and thus one finds for the anti-holomorphic differential algebra 
$\Omega_\cD^\hb\cA$
\be
\begin{array}{rcl}
\sd{0}{k} & : & \CL{0}{k} \lra \OD{0}{k}\\
          &   &                          \\
\sd{0}{k}(\etab_{\iad_1}\cdots\etab_{\iad_k}w^{\iad_1\cdots\iad_k}) &=&
\zb_{\iad_1}\cdots \zb_{\iad_k}w^{\iad_1\cdots\iad_k}\; ,
\end{array}
\ee
where $(\zb_{\iad_i})$ denotes the complex conjugate of $(z^{\ia_i})$.
The anti-holomorphic differential $d_\hb$ also is related to $d_h$ by
complex conjugation and is given as
\be
\begin{array}{rcl}
d_\hb & : &  \OD{0}{k} \lra \OD{0}{k+1}\\
 & & \\
d_\hb(\zb_{\iad_1}\cdots \zb_{\iad_k}w^{\iad_1\cdots\iad_k}) & = & {1\over k+1}
\zb_{\iad_1}\cdots \zb_{\iad_{k+1}}\left( \sum_{l=1}^{k+1}
\Db^{\iad_l} w^{\iad_1\cdots\iad_{l-1}\iad_{l+1}\cdots\iad_{k+1}}\right)\; .
\end{array}
\ee
Again it is
\be
d_\hb^2=0\; .
\ee

Having computed the purely holomorphic and anti-holomorphic part of $\ODA$
we now turn to the mixed forms of $\ODA$, i.e., to $\OD{k}{l}$ with $k\neq 0$
and $l\neq 0$. Thus we have to determine the correct product of holomorphic
and anti-holomorphic forms such that the total differential $d_\cD$ on 
$\ODA$ is nilpotent
\be
d_\cD^2 =0\; .
\ee
Since $d_\cD$ is determined by its action on holomorphic and anti-holomorphic
forms it is
\be
d_\cD = d_h + d_\hb
\ee
and hence the product of holomorphic and anti-holomorphic forms has to be
defined such that
\be
d_h d_\hb + d_\hb d_h = 0\; . \label{ho-aho}
\ee
However, since $D_\ia$ and $\Db_\iad$ do not anticommute eq.(\ref{ho-aho})
is not fulfilled for the product which is induced from $\CLA$. Thus there is
a non-trivial ideal in $\CLA$ which is generated by $\JD{1}{1}$.

$\JD{1}{1}$ itself is generated by elements of the form
$[\Db_\iad, a][D_\ia, b_+]$ and $[D_\ia, a^\prime][\Db_\iad, b_-]$ which obey
\be
a[D_\ia, b_+] = 0 \;\;\; , \;\;\; a^\prime[\Db_\iad, b_-] =0 
\;\;\; ,\;\;\; a,a^\prime \in \cA\; ,\;\; b_+\in \cA_+\;,\;\; b_-\in\cA_-
\ee 
At this point the $S$-numbers becomes important. This can already be seen
at first component of a superfield of the form $w=\pid(\delta \nu)$, 
$\nu \in \Omega^1\cA$ (we use the same labeling of the components of 
superfields as in the expansion in $\Th$, $\Thb$ of eq.(\ref{theta-exp})). 
For $\pid(\nu)=\eta^\ia a[D_\ia , b_+]$ we compute for the first component of 
$w$ in the $\Th$, $\Thb$ expansion
\be
\begin{array}{rcccl}
{w_1}_{\iad\ia}|_{S=2}&=&{a_3}_\iad{b_2}_\ia|_{S=2} &=&{v_3}_{\ia\iad}|_{S=2}\\
 & & & & \\
{w_1}_{\iad\ia}|_{S=-2}&=&{a_3}_\iad{b_2}_\ia|_{S=-2} &=&
{v_3}_{\ia\iad}|_{S=-2}\\
 & & & & \\
{w_1}_{\iad\ia}|_{S=0}&=&{a_3}_\iad{b_2}_\ia|_{S=0} &\neq&{v_3}_{\ia\iad}|_{S=0}
\end{array}\label{w1}
\ee
whereas the $S=0$ part of the 3rd component of $v$ in the $\Th$, $\Thb$
expansion is given as
\be
{v_3}_{\ia\iad}|_{S=0} = {a_3}_\iad {b_2}_\ia|_{S=0} 
-2i a_1\dslash_{\ia\iad} b_1\; . \label{v3}
\ee
Therefore we 
conclude that the first component of superfields in $\CL{1}{1}$ with $S=\pm 2$
are never in $\JD{1}{1}$. However, we also see from eq.(\ref{v3})
that there is for any ${w_1}_{\iad\ia}|_{S=0}$, given as
in eq.(\ref{w1}), an element $\nu\in\Omega^1\cA$ such that
\be
\pid(\nu) = 0 \;\;\; \hbox{and}\;\;\; 
\pid(\delta\nu)_1=\eta^\ia\etab_\iad {w_1}^\iad_\ia|_{S=0}\; 
.\label{allesweg}
\ee
Strictly speaking, for $\pid(\nu)=0$ it is not sufficient that $\pi(\nu)_3=0$.
However, it is straightforward to check that  
one can arrange $a$ and $b_+$ such that also all other components of 
$\pid(\nu)$ in the $\Th$, $\Thb$ expansion vanishes. 

Thus we have identified all elements of the form 
$a_1^{(i)}\dslash_{\ia\iad}b_1^{(i)}$ with 
$a_1^{(i)}, b_1^{(i)} \in C(\cM)$ as elements
of $\JD{1}{1}$.  Since $\cJ$ is an ideal we can multiply such elements
with arbitrary elements of $\cA$ and obtain
\be
\cN=\{\sum_i a^{(i)}\dslash_{\ia\iad}b^{(i)} , \;\; a^{(i)}, b^{(i)} \in \cA\}
\subseteq \JD{1}{1}\; . \label{j11}
\ee
Note, that if we had not generated the algebra $\cA$ by a spinor doublet which
are distinguished by the $S$-numbers, 
the space $\JD{1}{1}$ would be
the whole space $\CL{1}{1}$ and thus there would be no differential form 
with both holomorphic and anti-holomorphic indices. The reason for this is
that due to the $S$-numbers the range of the $\partial_ia$-part of $D_\ia$
is bigger then the range of the $i\dslash_{\ia\iad}\Thb^\iad$-part of $D_\ia$
which would not be the case if there is no split of components caused by
$S$-numbers.

What remains to be shown is that we have
determined all of $\JD{1}{1}$, i.e. that we can replace the "$\supseteq$"
by "$=$" in eq.(\ref{j11}). For this purpose it is convenient to define
a projection-operator $P_S^{(n)}$ which projects the components of the
$\Th$, $\Thb$-expansion of any superfield $w\in \CL{n-k}{k}$ onto the
parts with the highest $S$-numbers i.e.,  for $w\in\CL{k}{l}$ with
$k+l=n$, $k\neq 0$,  $l\neq 0$ and
\be
\begin{array}{rcccccl}
 & & & & |S(w_1)|&=&n  \\
 & & & & & & \\ 
 & &|S({w_2})| &=& |S({w_3})| &=& n+1 \\
 & & & & & &\\
|S({w_4})| &=& |S({w_5})| &= &|S(w_6)| &=& n + 2 \\
 & & & & & &\\
 & & |S({w_7})| &=& |S(w_8)| &=& n+3  \\
 & & & & & &\\
 & & & & |S({w_9})| &=& n+4
\end{array}\label{S-proj}
\ee
it is
\be
P_S^{(n)}(w) = w\label{S-proj1}
\ee
and $P_S^{(n)}(w)=0$ for all $w\in \CL{k}{l}$ which do not have components
with $S$-numbers as in eq(\ref{S-proj}). 
For later convenience we extend the definition
of $P_S$ to the holomorphic and anti-holomorphic part of $\CLA$ 
\be
P_S^{(n)} w = w\;\;\; ,\;\;\; \forall w\in(\CL{n}{0} + \CL{0}{n})\; .
\label{S-proj-ho}
\ee
Before we discuss the ideal generated by $\cN$ let us check that at the
level of 2-forms $\cN$ is the correct space by which one has to divide 
$\CL{1}{1}$ in order to obtain a well defined differential.
First we note that 
\be
\ker P_S^{(2)} \cap \CL{1}{1} = \cN\; .
\ee
Let $\pid(a^{(i)}\delta b^{(i)}) =v$ be an arbitrary 1-form. We compute for
the components of \linebreak
$P_S^{(2)}(\pid(\delta a^{(i)}\delta b^{(i)}))$ with a 
holomorphic and an anti-holomorphic index
\be
\begin{array}{cl}
 & P_S^{(2)}([D_\ia, a^{(i)}][\Db_\iad, b^{(i)}])+
P_S^{(2)}([\Db_\iad, a^{(i)}][ D_\ia, b^{(i)}])\\
 & \\ 
=& P_S^{(2)}([\partial_\ia, a^{(i)}][\parb_\iad, b^{(i)}])+
P_S^{(2)}([\parb_\iad, a^{(i)}][\partial_\ia, b^{(i)}])\\
 &  \\
 =& P_S^{(2)}([\partial_\ia, a^{(i)}[\parb_\iad, b^{(i)}]]_+) +
     P_S^{(2)}([\parb_\iad, a^{(i)}[\partial_\ia, b^{(i)}]]_+)\\
  & \\
 =& P_S^{(2)}([\partial_\ia, v ]_+) +
     P_S^{(2)}([\parb_\iad, v ]_+)\; .
\end{array}\label{N-ideal}
\ee
{}From this we conclude that if $v=0$ then it is
\be
P_S^{(2)}([D_\ia, a^{(i)}][\Db_\iad, b^{(i)}])+
P_S^{(2)}([\Db_\iad, a^{(i)}][ D_\ia, b^{(i)}]) = 0
\ee
which implies that
\be
\cN = \JD{1}{1}\; .
\ee
We now turn to the ideal which is generated by $\cN$.
With the definition of $P_S$ given in eq.(\ref{S-proj}) and 
eq.(\ref{S-proj-ho})it is straightforward to check that it is
\be
P_S^{(k+l+m+n)}(w_1 w_2) = P_S^{(k+l)}(w_1)  P_S^{(m+n)}(w_2)\;\;\; ,
\;\;\;\; w_1\in\CL{k}{l},\; w_2\in\CL{m}{n}
\ee
and
\be
P_S^{(k+l+2)}(w_1 w_2) = 0 \;\;\;\; ,\;\;\;\; w_1\in\CL{k}{l},\; 
w_2\in\cN\; .
\ee
Thus the ideal $I$ generated by $\cN$ is given by the kernels of the 
projectors $P^{(n)}_S$, $n\in\N$
\be
I= \bigcup_{n\geq 2} \ker P_S^{(n)}  \; .
\ee
{}From this it follows that product of holomorphic and anti-holomorphic forms 
are defined as follows: for any 
$v=z^{\ia_1}\cdots z^{\ia_k}v_{\ia_1\cdots\ia_k}\in \OD{k}{0}$ and
$w=\zb_{\iad_1}\cdots \zb_{\iad_l}w^{\iad_1\cdots\iad_l}\in \OD{0}{l}$ it is
\be
vw=z^{\ia_1}\cdots z^{\ia_k}\zb_{\iad_1}\cdots\zb_{\iad_l}
P_S^{(k+l)}(v_{\ia_1\cdots\ia_k}w^{\iad_1\cdots\iad_l})\; .
\ee
Thus we define the map $\sd{\bullet}{\bullet}$ from the super-Clifford algebra 
to the superdifferential algebra for any $k,l\in \N$
\be
\begin{array}{rcl}
\sd{k}{l} &:& \CL{k}{l}\lra \OD{k}{l}\\
 & & \\
\sd{k}{l}(\eta^{\ia_1}\cdots\eta^{\ia_k}\etab_{\iad_1}\cdots\etab_{\iad_l}
w_{\ia_1\cdots\ia_k}^{\iad_1\cdots\iad_l}) & = &
z^{\ia_1}\cdots z^{\ia_k}\zb_{\iad_1}\cdots\zb_{\iad_l}
P_S^{(k+l)}(w_{\ia_1\cdots\ia_k}^{\iad_1\cdots\iad_l})\; .
\end{array}
\ee
The extension of $d_h$ and $d_\hb$ to the mixed forms in $\OD{k}{l}$ is
obtained with the help of the projection $P_S$:
\be
\begin{array}{rclcl}
d_h & : & \OD{k}{l} \lra \OD{k+1}{l} & & \\
 & &  & &\\
d_h(w) & = & 
z^\ia P_S^{(k+l+1)}(D_\ia w - {(-1)}^{(k+l)} w D_\ia) &=&
z^\ia P_S^{(k+l+1)}([D_\ia , w ])
\end{array}
\ee
and
\be
\begin{array}{rclcl}
d_\hb & : & \OD{k}{l} \lra \OD{k}{l+1} & & \\
 & & & & \\
d_\hb(w) & = & 
z_\iad P_S^{(k+l+1)}(\Db^\iad w - {(-1)}^{(k+l)} w\Db^\iad) &=&
z_\iad P_S^{(k+l+1)}([\Db^\iad , w])\; .
\end{array}
\ee
The nilpotency of the holomorphic and anti-holomorphic differential is
again ensured by the symmetrization of the holomorphic and anti-holomorphic
indices. What remains to be checked is that
\be
d_h d_\hb + d_\hb d_h = 0\; .
\ee
However, this equation can be verified by the following computation: For
any $w\in \OD{k}{l}$ it is
\be
\begin{array}{rcl}
d_h(d_\hb w)&=&
\zb_\iad d_h (P_S^{(k+l+1)}(\Db^\iad w))=
z^\ia\zb_\iad P_S^{(k+l+2)}(D_\ia (P_S^{(k+l+1)}(\Db^\iad w)))\\
 & & \\
 &=& z^\ia\zb_\iad P_S^{(k+l+2)}(D_\ia \Db^\iad w)
\end{array}
\ee
and also
\be
\begin{array}{rcl}
d_\hb(d_h w)&=&
z^\ia d_\hb (P_S^{(k+l+1)}(D_\ia w))=
z^\ia\zb_\iad P_S^{(k+l+2)}(\Db^\iad (P_S^{(k+l+1)}(D_\ia w)))\\
 & & \\
 &=& z^\ia\zb_\iad P_S^{(k+l+2)}(\Db^\iad D_\ia w)\; .
\end{array}
\ee
Thus it is
\be
(d_h d_\hb + d_\hb d_h)(w) = 
z^\ia\zb_\iad P_S^{(k+l+2)}((D_\ia\Db^\iad + \Db^\iad D_\ia) w)=
z^\ia\zb^\iad P_S^{(k+l+2)}( 2i\dslash_{\ia\iad}w)=0\; .
\ee
{}From this it follows that $d_\cD=d_h+d_\hb$ is a nilpotent 
differential with $d_\cD^2=0$ on $n$-forms $n\in\N$.
This completes the construction of $\ODA$. 

Note, that the generalized differential algebra $\ODA$ itself does not 
contain any information of the underlying manifold $\cM$ in the sense that
the differential forms in $\ODA$ and the differential do not depend on
space-time derivatives. Although for the
construction of $\ODA$ the presence of $\cM$ played an important role, it 
turned out that all dependence on $C^\infty(\cM)$, via the $\dslash$-part 
of $\cD$, is contained in the differential ideal $\cJ$. Thus $\ODA$ is a
generalized differential algebra associated to the finite dimensional
Grassmann-algebra in $\cA$ which is multiplied by $C^\infty(\cM)$. 
For differential
forms in $\ODA$ which have both, holomorphic and anti-holomorphic indices,
this statement is a direct consequence of eq.(\ref{N-ideal}). 

For the pure holomorphic forms one can perform a change of coordinates
\be
x^\mu \lra y^\mu_- = x^\mu - i \Th \sigma^\mu\Thb\; .\label{Y-}
\ee
This induces the following transformation of the operators $D_\ia$ and 
$\Db_\iad$:
\be
\begin{array}{rcccl}
D_\ia & \lra & D_\ia^{(-)} &= & \partial_\ia\\
 & & & & \\
\Db_\iad & \lra & \Db_\iad^{(-)} &=& -\parb_\iad -
2i \Th^\ia\dslash_{\ia\iad}\; .
\end{array}\label{y-_trafo}
\ee
Since the pure holomorphic forms are built only out of commutators with
$D$ it follows from eq.(\ref{y-_trafo}) that they do not depend on
space-time derivatives.

For the pure anti-holomorphic forms there is a similar change of coordinates
\be
x^\mu \lra y^\mu_+ = x^\mu + i \Th \sigma^\mu\Thb\label{Y+}
\ee
which leads to
\be
\begin{array}{rcccl}
D_\ia & \lra & D_\ia^{(+)} &= & \partial_\ia +2i \dslash_{\ia\iad}\Thb\\
 & & & & \\
\Db_\iad & \lra & \Db_\iad^{(+)} &=& -\parb_\iad\; .
\end{array}\label{y+_trafo}
\ee
Thus we conclude that also the pure anti-holomorphic forms do not depend on
space-time derivatives.

Furthermore, we observe that differential forms with holomorphic and 
anti-holomorphic indices are invariant under transformations (\ref{Y-}),
(\ref{Y+}), i.e.
\be
P^{(k+l)}_S(\omega(x))=P^{(k+l)}_S(\omega(y_+))=P^{(k+l)}_S(\omega(y_-))
\;\;\; ,\;\;\; \forall \omega \in \CL{k}{l},\; k,l>0
\ee
which is a direct consequence of eq.(\ref{y-_trafo}) and eq.(\ref{y+_trafo}).

As a result of this discussion we may relate the generalized
differential algebra $\ODA$ to the algebra 
$C^\infty(\cM)\otimes\Lambda((\Th,\Thb)\times\C^2)$, where 
$\Lambda((\Th,\Thb)\times\C^2)$ denotes the $\Z_2$-graded analog
of the de Rham-algebra over the Grassmann-algebra generated by $(\Th,\Thb)\times\C^2$. Such $\Z_2$-graded de Rham algebras have already
been studied in the framework of noncommutative geometry in \cite{Rora} where
the relation between closed de Rham currents and cyclic cocycles over
a Grassmann algebra was established.

However, the algebra $\ODA$ is not isomorphic to 
$C^\infty\otimes\Lambda((\Th,\Thb)\times\C^2)$ because of the projection
operator $P_S$. The definition of $P_S$ in 
eqs.(\ref{S-proj},\ref{S-proj1},\ref{S-proj-ho}) can naturally be transferred
to $\Lambda((\Th,\Thb)\times\C^2)$. With this projection operator $P_S$
defined on $C^\infty\otimes\Lambda((\Th, \Thb)\times\C^2)$ (where $P_S$ is extended by the identity on $C^\infty(\cM)$) it is
\be
\ODA = P_S(C^\infty(\cM)\otimes \Lambda((\Th, \Thb)\times \C^2)\; .
\ee
Strictly speaking, the identification of pure holomorphic form and
pure anti-holomorphic form involves also coordinate transformations of
the form eq.(\ref{Y-}) and eq.(\ref{Y+}).
%18-apr-1996
\section{The inner product and supersymmetry transformations}
With the generalized differential algebra $\ODA$ we have all
necessary objects at hand to construct the covariant derivative and curvature,
the main objects in Yang-Mills theory.
However, what is still missing is an inner product on $\ODA$ which would
allow us to define an action. The standard procedure in noncommutative 
geometry uses the fact that there is a natural inner product on $\CLA$ 
which induces an inner product on $\ODA$ \cite{cobuch}. In principle we shall
also follow this construction although there will be some important 
deviations from the usual procedure.

Let us first define an inner product on $\CLA$. Therefore we recall that a
general element $\omega \in \CL{k}{l}$ is of the form
\be
\eta^{\ia_1}\cdots\eta^{\ia_k}\etab_{\iad_1}\cdots\etab_{\iad_l}
\otimes \omega^{\iad_1 \cdots\iad_l}_{\ia_1\cdots\ia_k}
\ee
where the first factor acts on $\cH_H$ and the second factor acts on $\cH_\pi$.
Using this notation, we define
\be
\cH_{(k,l)}=\{\omega^{\iad_1\cdots \iad_l}_{\ia_1\cdots\ia_k} |0\!> \; |\;
\omega \in \CL{k}{l}\}
\ee
which is completely analogous to the definition of $\cH_\cA$ in 
eq.(\ref{H_a-def}). Again this allows us to use the inner product on $\cH_\pi$
for the definition of a (degenerate) inner product on $\CL{k}{l}$ for any
$k,l \geq 0$
\be
<\!\omega | \nu\!> = 
\left((\omega^{\ia_1\cdots \ia_k}_{\iad_1\cdots\iad_l}|0\!>) ,
(\nu^{\iad_1\cdots\iad_l}_{\ia_1\cdots\ia_k}|0\!>)\right)\;\;\; ,\;\;\;
\omega,\;\nu \in \CL{k}{l}\; .\label{kl-prod}
\ee
This inner product is degenerate for the same reason as the inner
product (\ref{A-norm}) on $\cA$ is degenerate. However, on the subspaces
$(\CL{k}{l})_{v^{k+l-2m}}$, $2m\leq k+l$, the inner product is non-degenerate
if one of the components of $v\in\C^2$ is non zero. The natural inner product
on $\cA^{(k,l)}_0$ and the inner product defined in eq.(\ref{kl-prod}) are
related by
\be
\begin{array}{rcl}
<\! i_{v^{k+l-2m}}(\omega_0) , i_{v^{k+l-2n}}(\nu_0)\!> &=& 
{\scriptstyle
{(2(k+l+2-m-n))!\over ((k+l+2-m-n)!)^2}}(v_+v_-)^{k+l+2-m-n}(\omega_0 , \nu_0)\\
 & & \\
 &=& {\scriptstyle
{(2(k+l+2-m-n))!\over ((k+l+2-m-n)!)^2}}(v_+v_-)^{k+l+2-m-n}
\int_\cM (\omega_0^*\nu_0)|_{\Th\Th\Thb\Thb}\; , 
\end{array}
\ee
for all $\omega_0 \in (\CL{k}{l})_{v^{k+l-2m}}$ and for all 
$\nu_0 \in (\CL{k}{l})_{v^{k+l-2n}}$.
Since it is our aim to construct an action which is invariant under
supersymmetry transformation we are interested only in inner products on
the spaces $(\CL{k}{l})_{v^{k+l-2m}}$. For later convenience we rescale
these inner products
\be
\begin{array}{rl}
<\!\omega , \nu\!>_i = 
{\scriptstyle {((k+l+2-m-n)!)^2\over (2(k+l+2-m-n))!}}(v_+v_-)^{-k-l+m+n}
<\!\omega , \nu\!>,& \;\;
  \omega \in (\CL{k}{l})_{v^{k+l-2m}},\\
 & \\
 & \;\;\nu\in(\CL{k}{l})_{v^{k+l-2n}}
\end{array}
\ee
such that $i_{v^{k+l-2m}}$ becomes an isometric map.

We now come to the discussion about the relation of $\ODA$ and
supersymmetry transformations. Clearly we can transfer the embedding
$i_v$ of tensor superfield from $\CLA$ to $\ODA$, i.e. 
we define the subalgebra $(\ODA)_v\subset\ODA$
which carries a representation of the supersymmetry algebra for any
$k,l\in\N$ as
\be
(\OD{k}{l})_v = \sigma^{(k,l)}_\cD\circ i_v(\cA_0^{(k,l)})\; .
\ee
Note, that $\sigma^{(k,l)}_\cD$ is an invertible homomorphism from 
$(\CL{k}{l})_{v^{k+l}}$ to $(\OD{k}{l})_v$ if $ k,l>0$. Therefore we can 
define for any $k,l>0$
\be
c^{(k,l)}_v : (\OD{k}{l})_v \lra (\CL{k}{l})_{v^{k+l}}
\ee
as the inverse of $i_{v^{k+l}}$:
\be
c^{(k,l)}_v = {\sigma_\cD^{(k,l)}}^{-1}_{|(\CL{k}{l})_{v^{k+l}}}\; .
\ee
This map can be used to define an inner product on $(\ODA)_v$ which is
induced by the inner product on $(\CLA)_v$. However, the invariance under
supersymmetry transformations of this product is not automatically guaranteed.

For the pure holomorphic part of $\ODA$ we find that the image of $d_h$
acting on $(\ODA)_v$ is not contained in $(\ODA)_v$
\be
d_h (\OD{k}{0})_v \nsubset (\OD{\cdot}{0})_v\; .
\ee
The reason for this is the same as the one discussed at the end of section 3:
the $i\dslash_{\ia\iad}\Thb^\iad$-part of $d_h$ generates terms which are
not in $(\ODA)_v$. The same is true for the pur anti-holomorphic forms and
the differential $d_\hb$. 

The situation is different for forms with
mixed indices since here the disturbing part of
the derivative is projected out. Thus it is for all $k,l\in \N$ with $l>0$
\be
d_h \omega \in (\OD{k+1}{l})_v\;\;\; ,\;\;\;\forall \omega\in(\OD{k}{l})_v
\ee
and also for all $k,l\in\N$ with $k>0$
\be
d_\hb \omega \in (\OD{k}{l+1})_v\;\;\; ,\;\;\;\forall \omega\in(\OD{k}{l})_v\; .
\ee
However, supersymmetry transformations do not commute with the differentials
\be
\begin{array}{rcl}
{[d_h, (\Qbe)_v ]}\omega &=& i z^\ia\dslash_{\ia\iad}\veb^\iad_{(v)}\omega
\;\;\; ,\;\;\; \omega \in (\OD{k}{l})_v\; ,\; l>0\\
 & & \\
{[d_\hb, (\Qe)_v]}\omega &=& i \zb_\iad\dslab^{\iad\ia}{\ve_\ia}_{(v)}\omega
\;\;\; ,\;\;\; \forall \omega \in (\OD{k}{l})_v\; , \; k>0\; ,
\end{array}
\ee
where it is 
$({\ve_\ia}_{(v)} , {\veb^\iad}_{(v)}) = (\ve_\ia, \veb^\iad)\otimes v$.

On the other hand, it is for any $k,l>0$  and $ 0<2m \leq k+l$
\be
(\CL{k}{l})_{v^{k+l-2m}} \subset \JD{k}{l}\; .
\ee
Thus it is for any $\omega\in (\OD{k}{l})_v$
\be
d_h\omega 
= z^\ia P_S^{(k+l+1)}\left(\partial_\ia c_v^{(k,l)}(\omega)\right)
= z^\ia P_S^{(k+l+1)}\left(D_\ia c_v^{(k,l)}(\omega)\right)\; .
\ee
Although $\omega^\prime =\eta^\ia P_S^{(k+l+1)}(D_\ia c_v^{(k,l)}(\omega))$
also is not covariant under supersymmetry transformations, 
the product of $\omega^\prime$ with any other element in $(\OD{k}{l})_v$
with covariant transformation properties under supersymmetry transformations,
is invariant, i.e for any $\nu (\OD{k+1}{l})_v$ with 
$\nu=\sigma_\cD^{(k+1,l)}\circ i_{v^{k+l+1}}(\nu_0)$ and
$\omega=\sigma_\cD^{(k,l)}\circ i_{v^{k+l}}(\omega_0)$ it is
\be
\begin{array}{rcl}
<\!c_v^{(k+1,l)}(\nu), \omega^\prime\!>_i& =&
<\!c_v^{(k+1,l)}, \partial_\ia c_v^{(k,l)}(\omega) + 
i\dslash_{ia\iad}\Thb^\iad c_v^{(k,l)}(\omega)\!>_i\\
 & & \\
 &=& <\!c_v^{(k+1,l)}(\nu), i_{v^{k+l+1}}(\partial_\ia\omega_0) +
i_{v^{k+l-1}}(i\dslash_{\ia\iad}\Thb^\iad \omega_0)\!>_i\\
 & & \\
 &=& \int_\cM (\nu_0^* D_\ia \omega_0)|_{\Th\Th\Thb\Thb}
\end{array}\; .
\ee
The same arguments apply for the anti-holomorphic derivative, i.e. for
$d_h\omega\in (\OD{k}{l+1})_v$ the product 
$<\!c_v^{(k,l+1)}(\nu), \Db^\iad c_v^{(k,l)}\omega\!>_i$
is invariant under supersymmetry transformations for all 
$\nu \in (\OD{k}{l+1})_v$.
%18-apr-1996
\section{Supersymmetric Yang Mills theory}
Once the the generalized differential algebra $\ODA$ is known the covariant
derivative and curvature can be defined \cite{cobuch}. We repeat from this 
general procedure only the basic definitions which allows us to fix our 
notation. A comprehensive presentation of this topic can be found in 
\cite{cobuch,vabo,kabuch}.

The covariant derivative is defined with respect to some gauge group
which is in this framework
\be
\vG = \{ u\in\cA |\; uu^* =u^* u = 1\}\; .
\ee
There is a representation of this group on $\cH$, the Hilbert-space of the
spectral triple $(\cA, \cH, \cD)$ which is given by $\pi$, the represntation
of $\cA$. The operator
$\cD$ can be extended to a covariant derivative by adding a connection
1-form $A \in \Omega_\cD^1\cA \cong \pid(\Omega^1\cA)$, i.e. we define
the covariant derivative as an operator acting on $\cH$ by
\be
\nabsla = \cD + \sla{A}\; ,
\ee
where $\sla{A}\in\pid(\Omega^1\cA)$ is hermitean and obeys the following 
transforms rule
\be
\sla{A} \lra \sla{A}^\prime= u\sla{A}u^* + u\cD u^*\; .
\ee
The operator $\nabsla$ transforms covariant under gauge transformations
\be
\nabsla \lra {\nabsla\,}^{\prime} = u\nabsla u^*
\ee
Alternatively, the covariant derivative can be defined as an operator
acting on forms, i.e., as an operator acting on $\ODA\otimes\cH$
\be
\nabla = d_\cD + A \; ,
\ee
where $A=\sigma_\cD^1(\sla{A})\in\Omega_\cD^1\cA$ denotes the 1-form
corresponding to $\sla{A}$. Of course, $\nabla$ also transforms covariantly
under gauge transformations.

The curvature $F$ is defined as the square of the covariant derivative
\be
F = \nabla\nabla = d_\cD A + A A
\ee
and it is easy to show that also in the general framework of non-commutative
geometry this definition leads to a 2-form, i.e. $F\in \Omega^2_\cD\cA$,
which transforms covariantly under gauge transformations.

Let us now apply this general construction to the case where $\cA_n$ is the
tensor product of the commutative algebra of superfields as defined in sect.4
and the algebra of complex $n\times n$-matrices, $M_{n\times n}(\C)$, i.e.
\be
\cA_n = \cA \otimes M_{n\times n}(\C)\; .
\ee
The representation space $\cH$ has to be extended by a representation of
$M_{n\times n}$ such that it becomes a representation space $\cH_n$ of 
$\cA_n$. The only irreducible representation of the associative algebra
$M_{n\times n}(\C)$ is $\C^n$. Thus we take this irreducible representation
and obtain for $\cH_n$
\be
\cH_n = \cH \otimes \C^n\; .
\ee
The operator $\cD$ is extended trivially to an operator $\cD_n$
\be
\cD_n = \cD \otimes 1_{n\times n}
\ee
where $\cD$ is defined as in eq.(\ref{DO-def}).
 As a consequence of this setting the generalized 
differential forms in $\ODA$ become matrix-valued generalized 
differential-forms.

The gauge group $\vG$ is the group of superfields which are generated by the 
super-Lie algebra $\vg$
\be
\vg=\{ \Lambda \in\cA|\; \Lambda^*=\Lambda \}\;.
\ee
Thus any $u \in\vG$ can be written as $u=\exp(i\Lambda)$, $\Lambda\in\vg$.
Obviously the first component of any $u\in\vG$ of the 
$\Th$, $\Thb$-expansion is a bosonic $U(n)$-gauge-transformation. However,
any $u\in\vG$ represents a full superfield and therefore the bosonic
gauge group is extended by a nilpotent part, containing also Grassmann-odd
transformations.

We saw that the derivative $d_\cD$ of the $\ODA$, constructed in the previous
sections, splits into a holomorphic part $d_h$ and an anti-holomorphic part
$d_\hb$. Also the space of 1-forms $\Omega_\cD^1\cA$ can be decomposed into 
a holomorphic part $\OD{1}{0}$ and an anti-holomorphic part $\OD{0}{1}$.
Thus we can introduce the holomorphic and anti-holomorphic derivative
\be
\begin{array}{rcl}
\nabla &=& \nabla_h + \nabla_\hb \\
 & & \\
\nabla_h &=& d_h + A_h \\
 & & \\
\nabla_\hb &=& d_\hb + A_\hb\; ,
\end{array}
\ee
where $A_h=z^\ia A_\ia$, resp.~$A_\hb=\zb_\iad A^\iad$ denotes the holomorphic, resp.~anti-holomorphic
part of $A=A_h+A_\hb$.

This split propagates to the the 2-forms where we can decompose the
curvature as follows
\be
F= F_h + F_\hb + F_v
\ee
with
\be
\begin{array}{rcccl}
F_h &=& \nabla_h^2 &=& d_h A_h +A_h A_h \\
 & & & & \\
F_\hb &=& \nabla_\hb^2 &=& d_\hb A_\hb + A_\hb A_\hb
\end{array}
\ee
and
\be
F_v= \nabla_h\nabla_\hb +\nabla_\hb\nabla_h=d_h A_\hb + d_\hb A_h
+A_h A_\hb + A_\hb A_h\; .\label{F_V}
\ee
As in the usual approach to supersymmetric gauge-theory the full curvature
contains superfluous components \cite{WB} and one has to get rid of them
without spoiling covariance. The standard procedure is to impose the
constraint that all components of $F$ with 2 spinorial indices vanish.
In our case, this clearly would be to strong since it would
imply that the complete curvature vanishes. However, the standard constraints
in the usual approach have different reasonings: The requirement that the 
vectorial part of the curvature, i.e. $F_{\ia\iad}$ should vanish is simply
a redefinition of fields which is possible because of the presence of the
torsion term. This torsion term is absent in our approach. Therefore the
constraint $F_{\ia\iad}=0$ would be a real restriction and thus we drop this
constraint. 

The other constraints arise as a consequence of the chirality conditions
which reads
\be
\begin{array}{rcl}
\nabla_h \Phib &=& 0\\
 & & \\
\nabla_\hb \Phi& = & 0
\end{array}
\;\;\;\;\;\;\;\;\;\;\; ;\;\;\;\;\;\;\;\;\;\;\;\;\; \Phi , \Phib \in \cH_\pi\; .
\ee
These conditions can be applied consistently only if
\be
\begin{array}{rcccl}
\nabla_h\nabla_h &=& F_h&=& 0\\
 & & & & \\
\nabla_\hb\nabla_\hb &=& F_\hb &=& 0\; .
\end{array}
\ee 
This leads to the same
restrictions on $A$ as in the conventional approach. In components the constraints read
\be
\begin{array}{rcccl}
F_{\ia\ib} &=& D_\ia A_\ib  + D_\ib A_\ia + A_\ia A_\ib + A_\ib A_\ib &=&0 \\
 & & & &\\
F_{\iad\ibd} &=& \Db_\iad A_\ibd + \Db_\ibd A_\iad + A_\iad A_\ibd
+ A_\ibd A_\iad &=&0 \\
\end{array}\label{chircon}
\ee
The most general solution to the constraints in eq.(\ref{chircon}) are
\be
\begin{array}{rcl}
A_\ia &=& T^{-1}D_\ia T \\
 & & \\
A_\iad &=& S^{-1}\Db_\iad S \\
\end{array}
\ee
where $T,S\in\cA$ are general invertible superfields. They are related
by the requirement that $\nabsla$ is a self-adjoint operator. Thus it
is $\sla{A}^*=\sla{A}$ and hence ${A_h}_\ia=-{A_\hb}_\iad$. This implies
\be
S^*=T^{-1}\; .
\ee

Inserting this result in eq.(\ref{F_V}) we obtain for the remaining part of
the curvature
\be
\begin{array}{rcl}
F_{\ia\iad} & = &P^{2}_S(
D_\ia(T^*\Db_\iad {(T^{-1})}^*) + \Db_\iad(T^{-1}D_\ia T)\\
 & & \\
 & & +(T^*\Db_\iad {(T^{-1})}^*)(T^{-1}D_\ia T) +
(T^{-1}D_\ia T)(T^*\Db_\iad {(T^{-1})}^*))
\end{array}
\ee
which can be rewritten as
\be
\begin{array}{rcl}
F_{\ia\iad}&=& P^{(2)}_S\left(
T^* \Db_\iad(W^{-1}D_\ia W) {(T^{-1})}^*-
2iT^*\dslash_{\ia\iad}{(T^{-1})}^*\right)\\
 & & \\
 &=& P^{(2)}_S\left(\parb_\iad(W^{-1} \partial_\ia W)\right)\\
 & & \\
 &=& T^* W_{\ia\iad}{(T^{-1})}^*\; ,
\end{array}
\ee
where we have set $W=TT^*$ and
\be
W_{\ia\iad}= P^{(2)}_S\left(\Db_\iad (W^{-1} D_\ia W)\right)\; .\label{W-crum}
\ee
Comparing eq.(\ref{W-crum}) with supersymmetric Yang-Mills theory in the
chiral representation \cite{Gier} we see that
\be
W_{\ia}= \Db_\iad\Db^\iad (W^{-1} D_\ia W)
\ee
is the curvature in the usual approach to supersymmetric gauge theory if
$W\in\cA_0$. There it is only this quantity which transforms homogeneously.
Whereas it is straightforward to check, that in our framework $W_{\ia\iad}$
transforms homogeneously under chiral transformations $\Sigma$ 
with $\Db \Sigma =0$:
\be
\begin{array}{rcccl}
T &\lra & T^\prime &=& \Sigma^* T\\
  & & & & \\
W &\lra& W^\prime &=& \Sigma^* W \Sigma\\
  & & & & \\
W_{\ia\iad} &\lra& W^\prime_{\ia\iad} &=& \Sigma^{-1} W_{\ia\iad}\Sigma\; .
\end{array}
\ee
The reason for the homegenous transformation property of $W_{\ia\iad}$ is 
that the inhomgenous term which arises at the level of $\CL{1}{1}$ is 
in $\JD{1}{1}$.

This allows us to utilize the Wess-Zumino gauge \cite{WB} and to rewrite 
eq.(\ref{W-crum}) as
\be
W_{\ia\iad} = \Sigma^{-1} W^{WZ}_{\ia\iad} \Sigma\label{WZ}
\ee
with
\be
W^{WZ}_{\ia\iad}= \Db_\iad (\exp{-V_{WZ}}D_\ia \exp{V_{WZ}})
\ee
and
\be
V_{WZ}= -\Th\sigma^\mu\Thb A_\mu +i\Th\Th\Thb\chib - i\Thb\Thb\Th\chi 
+{1\over 2} \Th\Th\Thb\Thb D\; .\label{V_WZ}
\ee
Thus we infer that the curvature contains a vector-field, a Majorana spinor
and scalar field modulo chiral gauge transformations.

If it is $W\in \cA_v$ then it is $z^\ia\zb_\iad W_\ia^\iad \in {(\OD{1}{1})}_v$
and hence $z^\ia\zb_\iad F^\iad_\ia \in {(\OD{1}{1})}_v$. From eqs.(\ref{WZ},
\ref{V_WZ}) we conclude that the curvature is built out of a vector multiplet
modulo chiral gauge transformations. Since we want to
construct a supersymmetric invariant action we restrict ourselves to the case
$T \in \cA_v$ and hence $W\in \cA_v$. Furthermore, we can write
\be
T = i_v(T_0)\;\;\;\;\; , \;\;\;\;\; W = i_v(W_0)\; .
\ee
According to our discussion in the previous section a supersymmetric
invariant scalar $I$ for $F^2$ is given by
\be
\begin{array}{rcl}
I &=&tr(<\! T^*(\Db^\iad(W^{-1}D^\ia W))(T^{-1})^*,
T^*(\Db_\iad(W^{-1}D_\ia W))(T^{-1})^*\!>)\\ 
 & & \\
&= & tr(<\! \Db^\iad(W^{-1}D^\ia W),\Db_\iad(W^{-1}D_\ia W)\!>\\
 & & \\
 &=& tr\int_\cM 
(\Db^\iad(W_0^{-1}D^\ia W_0)\Db_\iad(W_0^{-1}D_\ia W_0))|_{\Th\Th\Thb\Thb}\\
 & & \\
 &=& -tr\int_\cM 
 ((W_0^{-1}D^\ia W_0)\Db\Db(W_0^{-1}D_\ia W_0))|_{\Th\Th\Thb\Thb}\\
 & & \\
 &=& -tr\int_\cM (\Db^2(W_0^{-1}D^\ia W_0)\Db^2(W_0^{-1}D_\ia W_0))|_{\Th\Th}
\end{array}\label{susy-act}
\ee
Inserting eqs.(\ref{V_WZ}) and (\ref{WZ}) in eq.(\ref{susy-act}) we obtain
\be
{\scriptstyle {1\over 16}}I= 
tr\int_\cM -F^{\mu\nu}F_{\mu\nu} -4i\chi\nabsla\chib +2 D^2 
+i\ve^{\mu\nu\lambda\rho}F_{\mu\nu}F_{\lambda\rho}
\ee
which is the action for supersymmetric Yang Mills theory \cite{WB}.
%14-apr-1996
\section{Conclusions}
In this article we have generalized the concept of the spectral triples to
algebras which contain both bosonic and fermionic degrees of freedom. The
unbounded selfadjoint operator of this triple was constructed out of the
spinorial generators of the supersymmetry algebra, i.e. the covariant
spinorial derivatives. The construction of the generalized differential
algebra out of this spectral triple was discussed in some detail. As a result
we obtained that one forms of this differential algebra are in the spin 
$1/ 2$-representations of the Lorentz-group and, more generally, that
$n$-forms are in the spin $n/ 2$-representations. This once more justifies
the well known notion that the covariant spinorial derivatives are the 
square-roots of the Dirac operator. 

For the resulting generalized differential algebra we found that only the 
finite dimensional structure of the Grassmann-algebra in $\cA$ is important, 
i.e. the generalized differential algebra itself does not contain more 
information about the underlying bosonic manifold $\cM$ then $C^\infty(\cM)$.
The bosonic part of the algebra becomes important when we consider 
supersymmetry which can be implemented only on a subalgebra of $\cA$ and
thus on a subalgebra of $\CLA$ and $\ODA$: the construction of an inner
product on $\ODA$ which is invariant under supersymmetry transformations 
involves space-time derivatives.

It is this inner product which marks important deviations from the standard
approach to Yang Mills theory in noncommutative geometry.
\begin{itemize}
\item[{\bf i)}] The representation of $\Omega\cA$ on $\cH$ allows to associate
to each element in $\CLA$ an element in $\cH$ and therefore the inner product 
on $\cH$ induces an inner product on $\CLA$. We did not use the Dixmier
trace for the definition of the inner product and thus we were not restricted 
to Euclidean space-time.
\item[{\bf ii)}] Since the inner product on $\CLA$ defined via the inner 
product on $\cH$ is indefinite on the subalgebra carrying a supersymmetry 
representation (and even degenerate on the whole algebra $\CLA$) we cannot
apply the standard procedure for the construction of an inner product on 
$\ODA$. Usually one identifies $\ODA$ as the orthogonal complement of the
ideal $\cJ$ in $\CLA$. This is not possible in our case since the inner
product on $\CLA$ is not positive definite. Therefore we had to use another
criterion to map elements of $\ODA$ into $\CLA$. For the subalgebra of
$\ODA$, which carries a representation of the supersymmetry algebra, 
we employed the requirement of invariance of the inner product under 
supersymmetry transformations.
\end{itemize}
Equipped with this inner product on the generalized differential algebra
we followed the standard procedure to construct Yang-Mills theory.
However, in our approach to Yang Mills theory we find as an immediate 
consequence of the relation between form degree and representation of the 
Lorentz-group that the curvature 2-form is a Lorentz-vector and therefore the 
lowest component in the $\Th, \Thb$-expansion of the curvature superfield is the 
vector-potential. The curvature superfield does not contain any space-time
derivative. Again it is the requirement of invariance under supersymmetry
which generates terms containing space-time derivatives in the action for
supersymmetric Yang-Mills theory.
%18-apr-1996


\newpage
\begin{thebibliography}{99}

\bibitem{cobuch}
A.~Connes.
\newblock {\em Noncommutative Geometry}.
\newblock Academic Press, (1994).

\bibitem{colo}
A.~Connes, J.~Lott.
\newblock {Particle} {Models} and {Noncommutative} {Geometry}.
\newblock {\em Nucl. Phys. {\bf B} Proc. Suppl.} {\bf 18B} (1990) 29--47.

\bibitem{cev}
R.~Coquereaux, G.~Esposito-Far\`ese, G.~Vaillant.
\newblock Higgs {Fields} as {Yang}--{Mills} {Fields} and {Discrete}
  {Symmetries}.
\newblock {\em Nucl. Phys. {\bf B}} {\bf 353} (1991) 689--706.\\
R.~Coquereaux.
\newblock {\em Higgs Fields and Superconnections, in `Differential Geometric
  Methods in Theoretical Physics' eds. C. Bartocci et al.}, {\em Lecture Notes
  in Physics} {\bf 375}, p. 3--24.
\newblock Springer-Verlag, (1991).

\bibitem{hps1}
R.~H\"au{\ss}ling, N.A.~Papadopoulos, F.~Scheck.
\newblock {$SU(2|1)$} symmetry, algebraic superconnections and a generalized
  theory of electroweak interactions.
\newblock {\em Phys. Lett. {\bf B}} {\bf 260} (1,2) (1991) 125--130.\\
R.~Coquereaux, G.~Esposito-Far\`ese, F.~Scheck.
\newblock Non {Commutative} {Geometry} and {Graded} {Algebras} in {Electroweak}
  {Interactions}.
\newblock {\em Int. J. of Mod. Phys. {\bf A}} {\bf 7} (26) (1992) 6555--6593.

\bibitem{hps2}
R.~H\"au{\ss}ling, N.A.~Papadopoulos, F.~Scheck.
\newblock Supersymmetry in the standard model of electroweak interactions.
\newblock {\em Phys. Lett. {\bf B}} {\bf 303} (3,4) (1993) 265--270.

\bibitem{dofrero}
S.~Doplicher, K.~Fredenhagen, J.E.~Roberts.
\newblock The {Quantum} {Structure} of {Spacetime} at the {Planck} {Scale} and
  {Quantum} {Fields}.
\newblock {\em Comm. Math. Phys.} {\bf 172} (1995) 187--220.

\bibitem{ma}
J.~Madore.
\newblock The fuzzy sphere.
\newblock {\em Class. Quant. Grav.} {\bf 9} (1992) 69--87.

\bibitem{mabu}
J.~Madore.
\newblock {\em Noncommutative Differential Geometry and its Physical
  Applications}, {\em London Math. Soc. Lecture Note Series} {\bf 206}.
\newblock Cambridge University Press, (1995).

\bibitem{groma}
H.~Grosse, J.~Madore.
\newblock A noncommutative version of the {Schwinger} model.
\newblock {\em Phys. Lett. {\bf B}} {\bf 283} (3,4) (1992) 218--222.

\bibitem{klim}
H.~Grosse, C.~Klim\v{c}\'{\i}k, P.~Pre\v{s}najder.
\newblock Simple {Field} {Theoretical} {Models} on {Noncommutative}
  {Manifolds}.
\newblock {\em prepr.} {\bf hep-th/9510177} (1995).

\bibitem{co}
A.~Connes.
\newblock Noncommutative geometry and reality.
\newblock {\em J. Math. Phys.} {\bf 36} (11) (1995) 6194--6231.

\bibitem{iokasch}
B.~Iochum, D.~Kastler, T.~Sch\"ucker.
\newblock Fuzzy {Mass} {Relations} in the {Standard} {Model}.
\newblock {\em prepr.} {\bf hep-th/9507150} (1995).

\bibitem{WB}
J.~Wess, J.~Bagger.
\newblock {\em Supersymmetry and Supergravity}.
\newblock Princeton Series in Physics. Princeton University Press, 2. edition,
  (1992).

\bibitem{cha}
A.H.~Chamseddine.
\newblock Connection between space--time supersymmetry and
  non--commutative geometry.
\newblock {\em Phys. Lett. {\bf B}} {\bf 332} (3,4) (1994) 349--357.

\bibitem{Jaffe}
A.~Jaffe.
\newblock Supersymmetric Analysis.
\newblock proceedings of Les Houches LXIV `Quantum Symmetries'.

\bibitem{vabo}
J.C.~V\'arilly, J.M.~Gracia-Bond\'{\i}a.
\newblock Connes' noncommutative differential geometry and the {Standard}
  {Model}.
\newblock {\em J. of Geom. and Phys.} {\bf 12} (1993) 223--301.

\bibitem{kabuch}
D.~Kastler, M.~Mebkhout. 
\newblock Lectures on non-commutative geometry and applications to 
  elementary particles. 
\newblock in preparation;\\
D.~Kastler, 
\newblock A detailed account of Alain Connes' version of the Standard Model 
  in non-commutative geometry Part I, Part II. 
\newblock {\em Rev.~Math.Phys.} {\bf 5} (1993) 477 and Part III 
  {\em Marseille prepr.} {\bf CPT-92/P.2824}.

\bibitem{Rora}R.~Coquereaux,E.~Ragoucy.
\newblock {\em Currents on Grassmann algebras}, {\em J. of Geom. and Phys.},
{\bf 15} (1995) 333-352.

\bibitem{Gier}
F.~Gieres.
\newblock {\em Geometry of Supersymmetric Gauge Theories}, {\em Lecture Notes
  in Physics} {\bf 302}.
\newblock Springer-Verlag, (1988).
\end{thebibliography}
\bye